\documentclass[a4paper]{article}
\usepackage{shadow,pifont,epsfig,wrapfig,times,graphics,color} 
\usepackage{latexsym}

\newcommand{\be}[1]{\begin{equation} \label{(#1)}}
\newcommand{\ee}{\end{equation}}
\newcommand{\ba}[1]{\begin{eqnarray} \label{(#1)}}
\newcommand{\ea}{\end{eqnarray}}

\begin{document}
\title{AN OVERVIEW OF THE HYPERCENTRAL CONSTITUENT QUARK
MODEL}  
\author{M.M. Giannini, E. Santopinto and A. Vassallo\\
{\small Dipartimento di Fisica dell'Universit\`a di Genova
      and I.N.F.N., Sezione di Genova}}

\maketitle                   

\abstract{We report on the recent results of the hypercentral Constituent
Quark Model (hCQM). The model contains a spin independent three-quark
interaction which is inspired by Lattice QCD calculations and reproduces
the average energy values of the $SU(6)$ multiplets.  The splittings are
obtained with a $SU(6)$-breaking interaction, which can include also an
isospin dependent term.
The model has been used for predictions concerning the electromagnetic
transition form factors giving a good description of the medium 
$Q^2$-behaviour. In particular the calculated helicity amplitude 
$A_{\frac{1}{2}}$ for the $S_{11}(1535)$ resonance agrees very well with
the recent CLAS data. 
Furthermore, we have shown for the first time that the decreasing of the
ratio of the
elastic form factors of the proton is due to relativistic
effects. Finally, the elastic nucleon form factors have been calculated
using a relativistic version of the hCQM and a relativistic quark
current.}

\section{Introduction}      
In recent years much attention has been devoted to the description of the 
internal nucleon structure in terms of quark degrees of freedom. Besides
the now classical Isgur-Karl model \cite{is}, the Constituent Quark Model
has been proposed in quite different approaches: the algebraic one
\cite{bil}, the hypercentral formulation \cite{pl} and the chiral model 
\cite{olof,ple}. In the following the hypercentral Constituent Quark Model
(hCQM), which has been used for a systematic calculation of various baryon
properties, will be briefly reviewed.

\section{The hypercentral model}
The experimental $4-$ and $3-$star non strange resonances can be arranged
in  $SU(6)-$multiplets (see Fig.~1). This means that the quark dynamics
has a dominant $SU(6)-$invariant part, which accounts for the average
multiplet energies. In the hCQM it is assumed to be
\cite{pl}
\begin{equation}\label{eq:pot}
V(x)= -\frac{\tau}{x}~+~\alpha x,
\end{equation}
where $x$ is the hyperradius 
\begin{equation}
x=\sqrt{{\vec{\rho}}^2+{\vec{\lambda}}^2} ~~,
\end{equation}
where $\vec{\rho}$ and $\vec{\lambda}$ are the Jacobi coordinates
describing the
internal quark motion. The dependence of the potential on the hyperrangle
$\xi=arctg(\frac{{\rho}}{{\lambda}})$ has been neglected.

\begin{figure}[ht]
\begin{center}
\includegraphics[width=14cm]{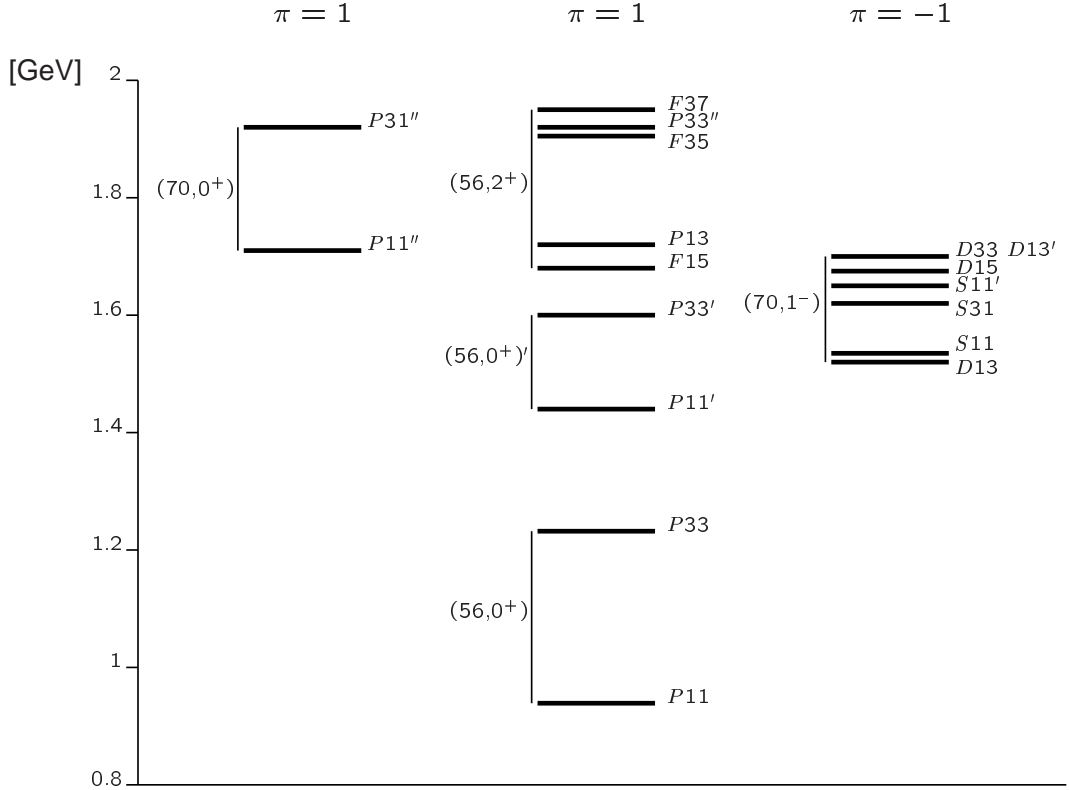}
\end{center}
\caption{\small The experimental spectrum of the 4- and 3-star non strange
resonances. On the left the standard assignements to $SU(6)$ multiplets
is reported, with the total orbital angular momentum and the parity.}
\end{figure}

Interactions of the type linear plus Coulomb-like have been used since
time for the meson sector, e.g. the Cornell potential. This form has been
supported by recent Lattice QCD calculations \cite{bali}. In the case of
baryons a so called hypercentral approximation has been
introduced \cite{has,rich}, which amounts to average any two-body 
potential for the three quark system over the hyperangle $\xi$ and works
quite well, specially for the lower part of the spectrum \cite{hca}. In
this respect, the hypercentral potential Eq.\ref{eq:pot} can be considered
as the hypercentral approximation of the Lattice QCD potential. On the
other hand, the hyperradius $x$ is a collective coordinate and therefore
the hypercentral potential contains also three-body effects.

The hypercoulomb term $1/x$ has important
features \cite{pl,sig}: it can be solved analytically and the resulting
form factors have a power-law behaviour, at variance with the widely used
harmonic oscillator; moreover, the negative parity states are exactly
degenerate with the first positive parity excitation, providing a good
starting point for the description of the spectrum.
 
The splittings within the multiplets are produced by a perturbative term
breaking $SU(6)$, which as a first approximation can be assumed to be the
standard hyperfine interaction $H_{hyp}$ \cite{is}.
The three quark hamiltonian for the hCQM is then:
\begin{equation}\label{eq:ham}
H = \frac{p_{\lambda}^2}{2m}+\frac{p_{\rho}^2}{2m}-\frac{\tau}{x}~
+~\alpha x+H_{hyp},
\end{equation}
where $m$ is the quark mass (taken equal to $1/3$ of the nucleon mass). 
The strength of the hyperfine interaction is determined in order to
reproduce the $\Delta-N$ mass difference, the remaining two free
parameters are fitted to the spectrum, reported in Fig. 2, leading to the
following values:
\begin{equation}\label{eq:par}
\alpha= 1.16~fm^{-2},~~~~\tau=4.59~.
\end{equation}

\begin{figure}[ht]
\begin{center}
\includegraphics[width=12cm]{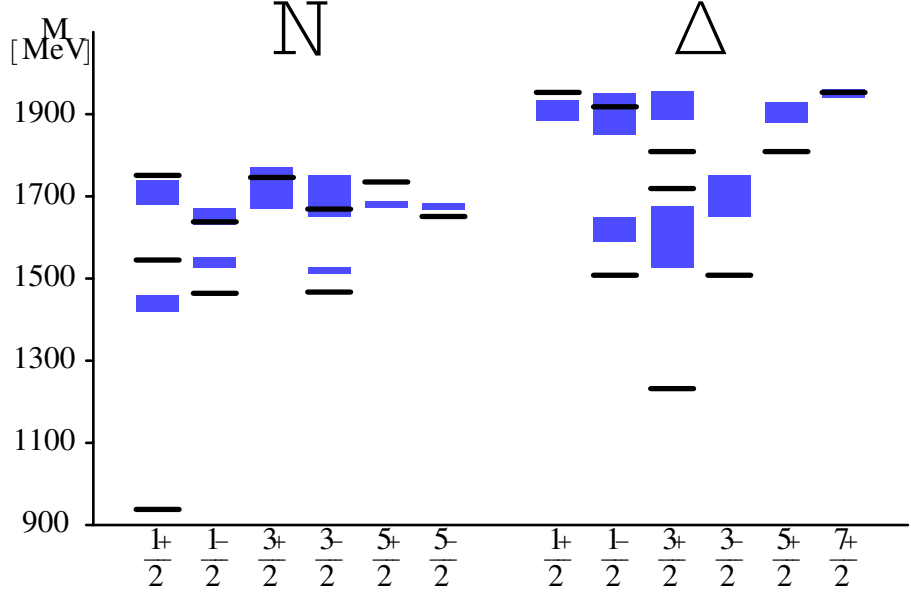}
\end{center}
\caption{\small The spectrum obtained with the hypercentral model Eq.
(3) and the parameters Eq. (4) (full lines)), compared with the
experimental data of PDG \cite{pdg} (grey boxes).}
\end{figure}

Keeping these parameters fixed, the model has been applied to calculate 
various physical quantities of interest: the photocouplings \cite{aie},
the
electromagnetic transition amplitudes \cite{aie2}, the elastic nucleon
form factors \cite{mds} and the ratio between the electric and magnetic
proton form factors \cite{rap}. Some results of such parameter free
calculations are presented in the next section.

\section{The results}

The electromagnetic transition amplitudes, 
$A_{1/2}$ and $A_{3/2}$, are defined as the matrix elements of
the  transverse electromagnetic interaction, $H_{e.m.}^t$, between the 
nucleon, $N$, and the resonance, $B$, states:
\begin{equation}\label{eq:amp}
\begin{array}{rcl}
A_{1/2}&=& \langle B, J', J'_{z}=\frac{1}{2}\ | H_{em}^t| N, J~=~
\frac{1}{2}, J_{z}= -\frac{1}{2}\
\rangle\\
& & \\
A_{3/2}&=& \langle B, J', J'_{z}=\frac{3}{2}\ | H_{em}^t| N, J~=~
\frac{1}{2}, J_{z}= \frac{1}{2}\
\rangle\\\end{array}
\end{equation}
The transition operator is assumed to be
\begin{equation}\label{eq:htm}
H^t_{em}~=~-~\sum_{j=1}^3~\left[\frac{e_j}{2m_j}~(\vec{p_j} \cdot
\vec{A_j}~+
~\vec{A_j} \cdot \vec{p_j})~+~2 \mu_j~\vec{s_j} \cdot (\vec{\nabla} 
\times \vec{A_j})\right]~~,
\end{equation}
where spin-orbit and higher order corrections are neglected 
\cite{cko, ki}. In 
Eq. (\ref{eq:htm}) $~~m_j$, $e_j$, $\vec{s_j}$ , $\vec{p_j}$ and 
$\mu_j~=~\frac{ge_j}{2m_j}$ denote the mass, the electric charge, the
spin, the momentum and the magnetic moment of the j-th quark,
respectively, and $\vec{A_j}~=~\vec{A_j}(\vec{r_j})$
is the photon field; quarks are assumed to be pointlike.

The proton photocouplings of the hCQM \cite{aie} (Eq.~(\ref{eq:amp})
calculated  at the photon point), in comparison with other
calculations \cite{bil,ki,cap}, have the same overall behaviour,
having the same SU(6)
structure in common, but in many cases they all show a lack of strength.

\begin{figure}[ht]
\begin{center}
\includegraphics[width=10cm,angle=90]{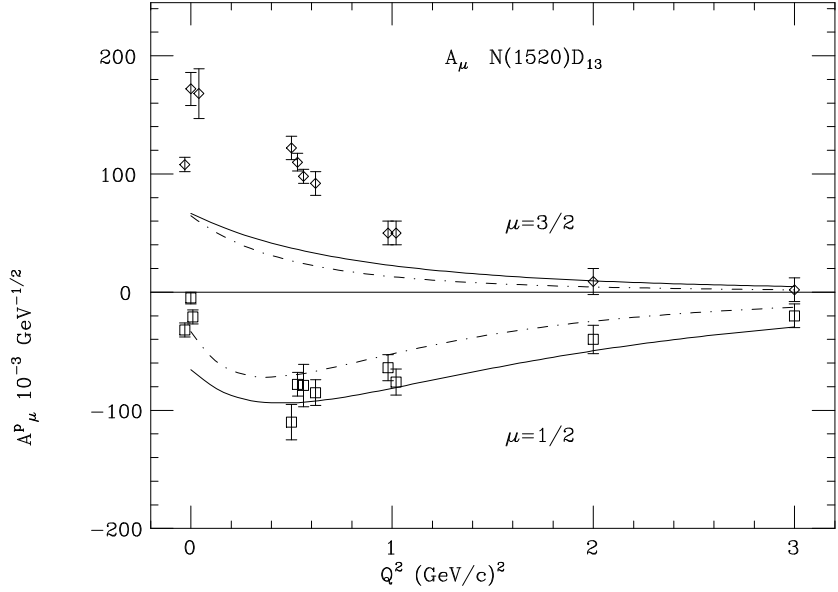}
\end{center}
\caption{The helicity amplitudes for the $D_{13}(1520)$ resonance,
calculated with the hCQM of Eqs. (3) and (4) and the electromagnetic
transition operator of Eq. (6) (full curve, \cite{aie2}).
The dashed curve is obtained with the analytical version of the hCQM
(\cite{sig}), where the behaviour of the quark wave function is determined
mainly by the hypercoulomb potential. The data are from the compilation of
ref. \cite{burk}}
\end{figure}

\begin{figure}[ht]
\begin{center}
\includegraphics[width=14cm]{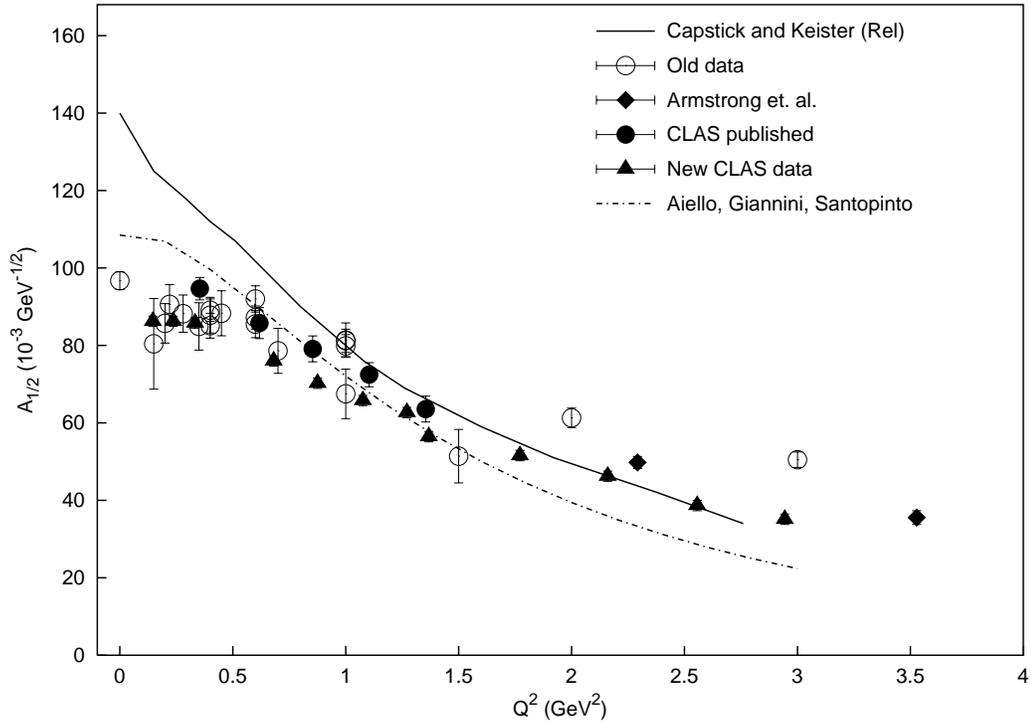}
\end{center}
\caption{\small The helicity amplitude for the $S_{11}(1535)$ resonance,
calculated with the hCQM of Eqs. (3) and (4) and the electromagnetic
transition operator of Eq. (6) (dashed curve, \cite{aie2}) and the model of
ref.\cite{cap}. The data are taken from the compilation of ref.
\cite{burk2}}
\end{figure}

Taking into account the $Q^2-$behaviour of the transition matrix elements 
of Eq.~(\ref{eq:amp}), one can calculate the hCQM helicity amplitudes in
the Breit frame \cite{aie2}. The hCQM results for the $D_{13}(1520)$ and
the $S_{11}(1535)$ resonances \cite{aie2} are given in Fig. 3 and 4,
respectively. The agreement in the case of the $S_{11}$ is remarkable, the
more so since the hCQM curve has been published three years in advance
with respect to the recent TJNAF data \cite{dytman}.

In general the $Q^2$-behaviour is reproduced, except for
discrepancies at small $Q^2$, especially in the
$A^{p}_{3/2}$ amplitude of the transition to the $D_{13}(1520)$ state. 
These discrepancies, as the ones observed in the photocouplings, can be
ascribed either to the non-relativistic character of
the model or to the lack of explicit quark-antiquark configurations, 
which may be important at low $Q^{2}$ .  The kinematical relativistic
corrections at the level of boosting the nucleon  and the resonance
states to a common frame are not  responsible for these discrepancies,  as
we have demonstrated in Ref.~\cite{mds2}. Similar results are obtained for 
the other negative parity resonances \cite{aie2}. 
It should be mentioned that the r.m.s. radius of the proton corresponding 
to the parameters of Eq.~(\ref{eq:par}) is $0.48~fm$, which is just the
value obtained in \cite{cko} in order to reproduce the $D_{13}$
photocoupling. Therefore the missing strength at low $Q^2$ can be ascribed
to the lack of quark-antiquark effects, probably more important in the
outer region of the nucleon.

\section{The isospin dependence}

The well known Guersey-Radicati mass formula \cite{gura}
contains a flavour dependent term, which is essential for the description
of the strange baryon spectrum.
In the chiral Constituent Quark Model \cite{olof,ple}, the non
confining part of the   
potential is provided by the interaction with the Goldstone bosons,
giving rise to a spin- and flavour-dependent part, which is crucial in
this approach for the description of the lower part of the spectrum.
More generally, one can expect that the quark-antiquark pair production 
can lead to an effective residual quark interaction containing an isospin
(flavour) dependent term.

\begin{figure}[ht]
\begin{center}
\includegraphics[width=12cm]{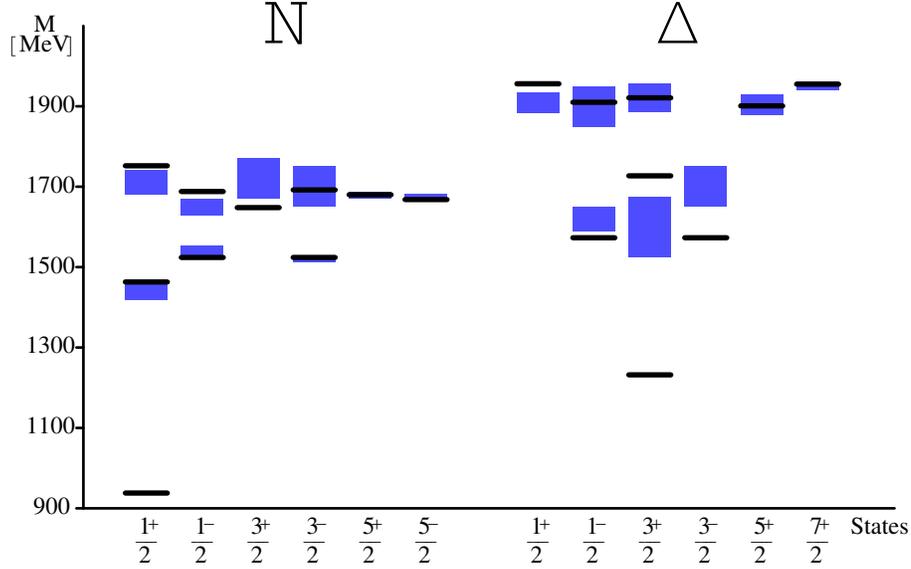}
\end{center}
\caption{\small The spectrum obtained with the hypercentral model
containing isospin dependent terms Eq. (7) \cite{iso} (full lines)),
compared with the
experimental data of PDG \cite{pdg} (grey boxes).}
\end{figure}

Therefore, we have introduced isospin dependent terms in the hCQM 
hamiltonian. The complete interaction used is given by \cite{iso}
\begin{equation}\label{tot}
H_{int}~=~V(x) +H_{\mathrm{S}} +H_{\mathrm{I}} +H_{\mathrm{SI}}~,
\end{equation}
where $V(x)$ is the linear plus hypercoulomb SU(6)-invariant potential of Eq.
\ref{eq:pot}, while the remaining terms are the residual SU(6)-breaking
interaction, responsible for the splittings within the multiplets. 
${H}_{\mathrm{S}}$ is a smeared standard hyperfine term,  
${H}_{\mathrm{I}}$ is isospin dependent and  ${H}_{\mathrm{SI}}$ 
spin-isospin dependent.
The resulting spectrum for the 3*- and 4*- resonances is shown in Fig.~5
\cite{iso}. The contribution of the
hyperfine interaction to the $N-\Delta$ mass difference is only about
$35\%$, while the remaining splitting comes from the
spin-isospin term, $(50\%)$, and from the isospin one, $(15\%)$.
It should be noted that the position of the Roper and the negative 
parity states is well reproduced.

\section{Relativity}

The relativistic effects that one can introduce starting from a non
relativistic quark model are:
a) the relativistic kinetic energy;
b) the boosts from the rest frames of the initial and final baryon to a 
common (say the Breit) frame;
c) a relativistic quark current. All these features are not equivalent to 
a fully relativistic dynamics, which is still beyond the present 
capabilities of the various models.

The potential of Eq.\ref{eq:pot} has been refitted using the correct 
relativistic kinetic energy
\begin{equation}\label{eq:hrel}
H_{rel} = \sum_{i<j} \sqrt{p_{ij}^2+m^2}-\frac{\tau}{x}~
+~\alpha x+H_{hyp}.
\end{equation}
 The resulting spectrum is not much different from the non relativistic
one and the parameters of the potential are only slightly modified. 

\begin{figure}[ht]
\begin{center}
\includegraphics[width=12cm]{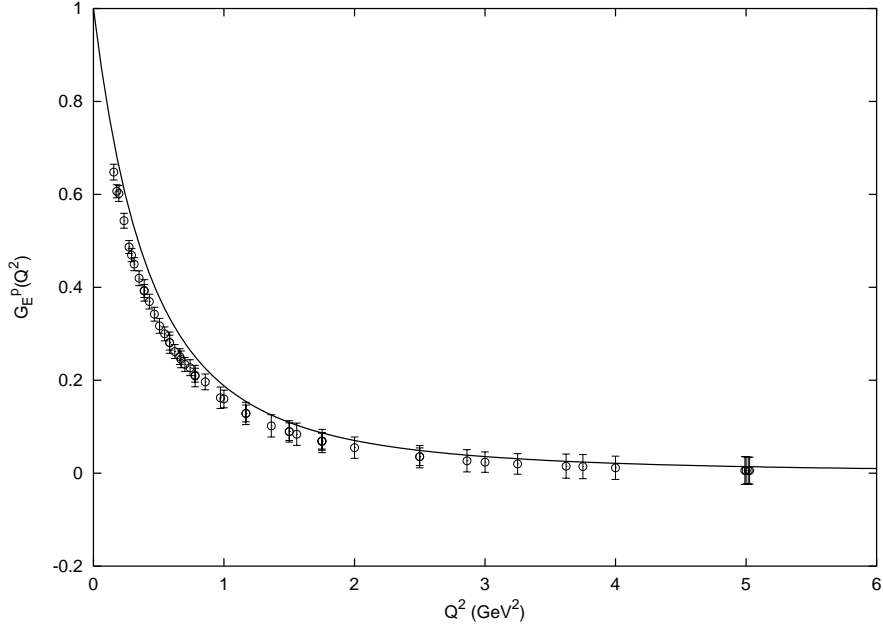}
\end{center}
\caption{\small The electric proton form
factor, calculated with the relativistic hCQM of Eq. (8) and a
relativistic quark current \cite{mds3}.}
\end{figure}

\begin{figure}[ht]
\begin{center}
\includegraphics[width=12cm]{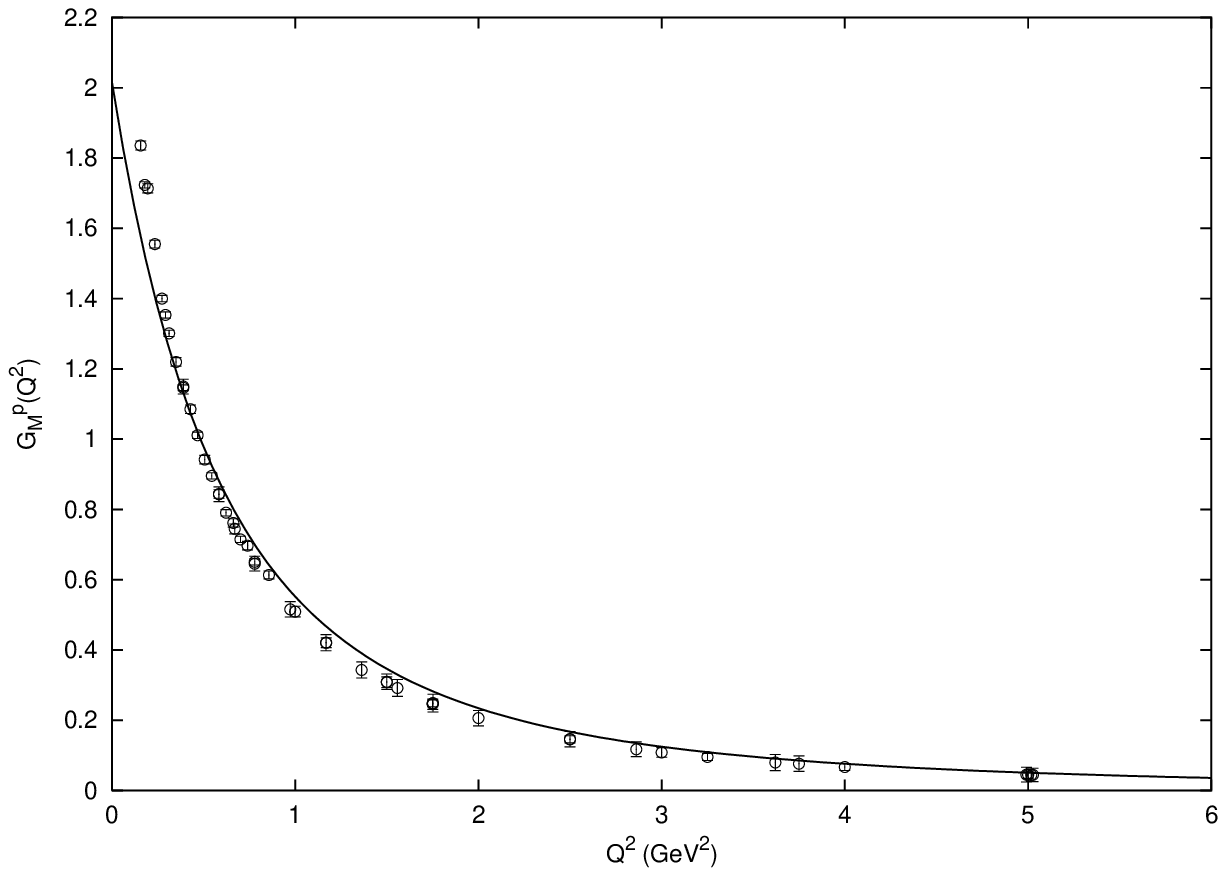}
\end{center}
\caption{\small The magnetic proton form
factor, calculated with the relativistic hCQM of Eq. (8) and a
relativistic quark current \cite{mds3}.}
\end{figure}

\begin{figure}[ht]
\begin{center}
\includegraphics[width=12cm]{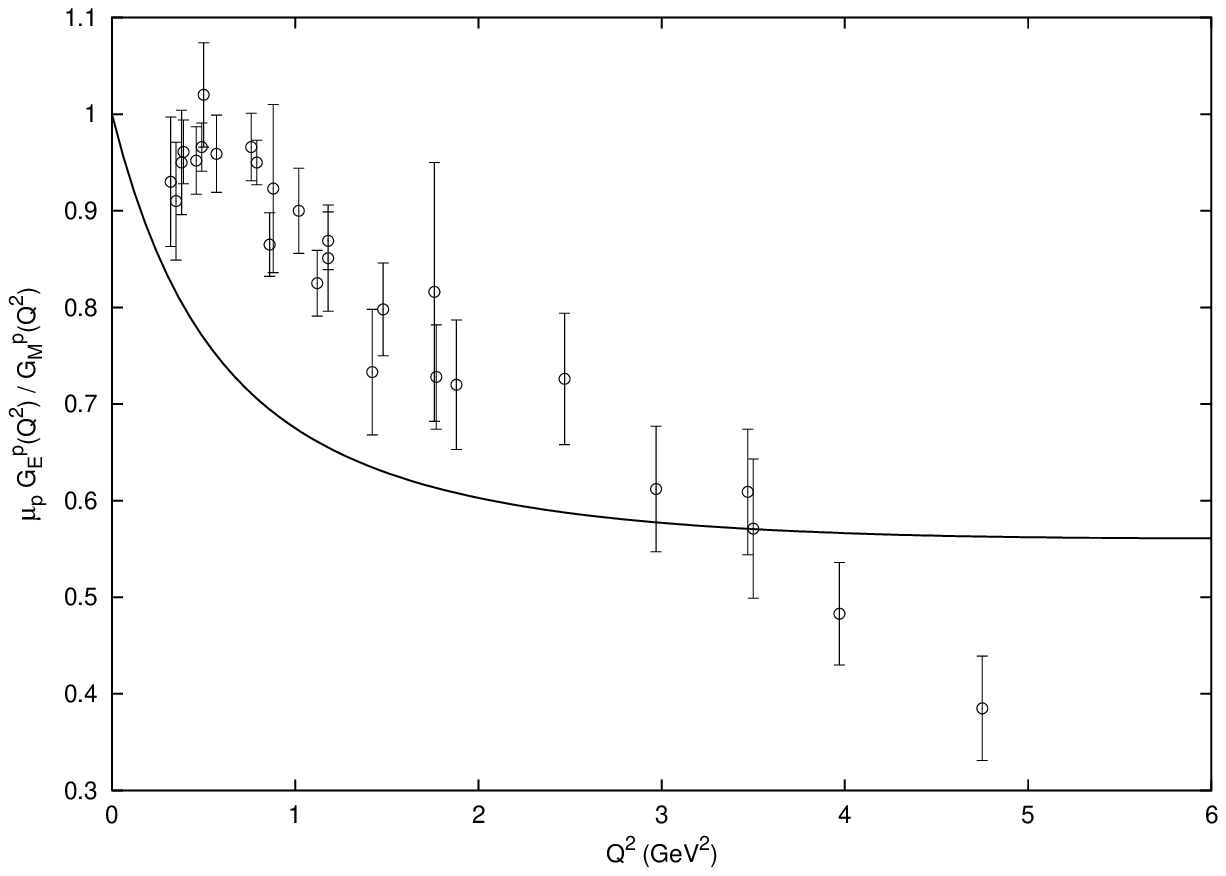}
\end{center}
\caption{\small The ratio between the electric and magnetic proton form
factors, calculated with the relativistic hCQM of eq. (8) and
a relativistic current \cite{mds3},
compared with the TJNAF data \cite{ped,gay}.}
\end{figure}

The boosts and a relativistic quark current expanded up to lowest order 
in the quark momenta has been used both for the elastic form factors of
the nucleon \cite{mds} and the helicity amplitudes \cite{mds2}. In the
latter case, as already mentioned, the relativistic effects are quite
small and do not alter the agreement with data discussed previously. For
the elastic form factors, the relativistic effects are quite strong and
bring the theoretical curves much closer to the data; in any case they are
responsible for the decrease of the ratio between the electric and magnetic
proton form factors, as it has been shown for the first time in Ref.~ 
\cite{rap}, in qualitative agreement with the recent 
Jlab data \cite{ped}.

A relativistic quark current, with no expansion in the quark momenta, and 
the boosts to the Breit frame have been applied to the calculation of the
elastic form factors in the relativistic version of the hCQM
Eq. (\ref{eq:hrel}) \cite{mds3}.
The resulting theoretical form factors of the proton, calculated, it
should be stressed, without free parameters and assuming pointlike quarks,
are good (see Figs. ~6 and 7), with some discrepancies at low $Q^2$,
which, as discussed
previously, can be attributed to the lacking of the quark-antiquark pair 
effects. The corresponding ratio between the electric and magnetic proton
form factors is given in
Fig. 8: the deviation from unity reaches almost the $50\%$ level, not far
from the new TJNAF data \cite{gay}.

\section{Conclusions}

The hCQM is a generalization to the baryon sector of the widely used
quark-antiquark
potential containing a coulomb plus a linear confining term. The three free
parameters have been adjusted to fit the spectrum \cite{pl} and then the 
model has been used for a systematic calculation of various physical
quantities: the photocouplings \cite{aie}, the helicity amplitudes for the
electromagnetic excitation of negative parity baryon resonances
\cite{aie2,mds2}, the elastic form factors of the nucleon \cite{mds,mds3}
and the ratio between the electric and magnetic proton form
factors \cite{rap,mds3}. The agreement with data is quite good, specially
for the helicity amplitudes, which are reproduced in the medium-high $Q^2$
behaviour, leaving some discrepancies at low (or zero) $Q^2$, where the
lacking quark-antiquark contributions are expected to be effective. It
should be noted that the hypercoulomb term in the potential is the main
responsible of such an agreement \cite{sig}, while for the
spectrum a further fundamental aspect is provided by the isospin dependent
interactions \cite{iso}.

\end{document}